\documentclass{optica-article}

\journal{opticajournal} % for journals or Optica Open

\articletype{Research Article}

\begin{document}

\title{Dynamic dark-field FFOCT and dynamic reflection differential phase contrast for label-free functional imaging at reflective biomaterial interfaces}

\author{Tual Monfort\authormark{1,*}}

\address{\authormark{1}Sorbonne Université, INSERM, CNRS, Institut de la Vision, 17 rue Moreau, F-75012, Paris, France\\
%\authormark{2}Publications Department, Optica Publishing Group, 2010 Massachusetts Avenue NW, Washington, DC 20036, USA\\
}

\email{\authormark{*}tual.monfort@inserm.fr} %% email address is required; see note below about the corresponding author designation

% use {asbstract*} to suppress the copyright line. Copyright information will be added in production

\begin{abstract*}

Strong reflections from metallic and engineered substrates severely limit label-free functional imaging of living cells at biomaterial interfaces, neural electrodes, and implantable devices. Here we introduce two complementary approaches for recovering intracellular dynamic contrast at highly reflective interfaces. Dynamic dark-field full-field optical coherence tomography (D-dFFOCT) suppresses the dominant substrate reflection and restores intracellular visibility through selective detection of scattered light. In parallel, asymmetric illumination generates a distinct directional dynamic contrast that is most consistently interpreted as dynamic reflection differential phase contrast (D-RDPC). Both approaches reveal intracellular activity that remains poorly visible with conventional dynamic full-field optical coherence tomography. D-RDPC exhibits characteristic signatures of phase-gradient imaging, including contrast reversal upon illumination inversion, enhancement with increasing illumination asymmetry, and recovery of spatial localization through directional Hilbert-transform reconstruction. Together, these results establish new strategies for functional imaging at reflective interfaces and suggest that differential phase contrast signals can support temporal fluctuation analysis.

\end{abstract*}

%%%%%%%%%%%%%%%%%%%%%%%%%%  body  %%%%%%%%%%%%%%%%%%%%%%%%%%

\section{Introduction}

The interaction between living cells and artificial surfaces is central to a wide range of biomedical technologies, including implantable devices, biomaterials, neural interfaces, biosensors, and engineered cell-culture platforms \cite{anderson2008foreign,williams2008mechanisms,franz2011immune,grainger2013all,sridharan2015biomaterial}. Understanding these interactions requires imaging techniques capable of probing cellular organization and intracellular activity directly at the cell--material interface while preserving physiological conditions \cite{zhang2018imaging}. Although fluorescence microscopy provides powerful molecular specificity, the use of exogenous labels can perturb cellular behavior, complicate long-term studies, and remains challenging to translate to many implant-related and clinical imaging scenarios \cite{progatzky2013seeing,park2018quantitative}. Label-free optical imaging therefore represents an attractive route for studying living systems at biomaterial interfaces.

Full-field optical coherence tomography (FF-OCT) is a label-free interferometric imaging modality capable of producing en face tomographic images with subcellular spatial resolution and high acquisition speed \cite{dubois2002high,dubois2004full}. Dynamic full-field optical coherence tomography (D-FFOCT) further extends these capabilities by exploiting temporal fluctuations of the interferometric signal to reveal intracellular activity in living tissues and cell cultures \cite{apelian2016dynamic,thouvenin2017dynamic}. This dynamic contrast has been associated with organelle motility, intracellular transport, membrane fluctuations, and refractive index redistribution, providing endogenous functional information without the need for exogenous markers \cite{apelian2016dynamic,scholler2020dynamic}. D-FFOCT has demonstrated promising capabilities in applications ranging from cellular phenotyping and tissue characterization to retinal imaging and histopathology \cite{scholler2019proliferative,defienne2020dynamic,groux2022retinal,monfort2023dynamic}.

Many applications of interest, however, involve cells cultured on highly reflective substrates. Metallic implants, reflective biomaterials, microelectrode arrays, coated surfaces, and engineered interfaces generate strong specular reflections that can dominate the detected optical signal \cite{anderson2008foreign,williams2008mechanisms,franz2011immune,drexler2008oct}. In our observations, conventional D-FFOCT exhibits a marked reduction of intracellular contrast under such conditions, severely limiting its applicability to the study of cell--surface interactions. While interface self-referenced D-FFOCT successfully suppresses fringe artefacts associated with weakly reflective interfaces \cite{monfort2023isr}, strong specular reflections remain a major challenge and can substantially reduce the visibility of intracellular dynamics \cite{Monfort2026}. Despite the practical importance of this problem, potential solutions remain largely unexplored.

In parallel with the development of dynamic interferometric microscopy, differential phase contrast (DPC) imaging has become one of the most widely used label-free microscopy modalities for visualizing transparent biological specimens \cite{hoffman1975modulation,mehta2009quantitative,tian2015computational}. By introducing controlled illumination asymmetry, DPC converts phase-gradient information into measurable intensity variations and provides strong contrast from weak phase objects. More recently, reflection-mode implementations have demonstrated that DPC-like contrast can also be generated in epi-illumination geometries through the interaction of asymmetric illumination with forward- and backward-scattered fields \cite{ford2012oblique,ford2014sobm,matlock2020inverse,ledwig2021quantitative,ma2022reflection}. These developments have remained primarily focused on structural imaging, and whether phase-gradient contrast can be exploited as a source of endogenous dynamic information remains unknown.

Here, we investigate functional imaging at reflective interfaces using a modified FF-OCT platform. We demonstrate that intracellular contrast can be recovered through two complementary approaches. First, we introduce dynamic dark-field FFOCT (D-dFFOCT), which suppresses the dominant specular reflection and restores intracellular dynamic contrast through selective detection of scattered light. Second, we show that asymmetric illumination generates a distinct form of directional dynamic contrast that is most consistently interpreted as dynamic reflection differential phase contrast (D-RDPC). Through experimental observations and Fourier-optics analysis, we demonstrate that the observed signal exhibits the characteristic signatures of a DPC imaging mechanism while encoding biologically meaningful temporal fluctuations.

To the best of our knowledge, this work constitutes the first demonstration of both D-dFFOCT and D-RDPC. Beyond enabling functional imaging at reflective interfaces, our results suggest that DPC microscopy itself may be extended from a structural modality toward functional imaging through temporal fluctuation analysis, creating new opportunities for label-free imaging of living systems.

\section{Results}

\subsection{D-dFFOCT and D-RDPC restore intracellular dynamic contrast at reflective interfaces}

\begin{figure}[htbp]
\centering
\includegraphics[width=1.0\textwidth]{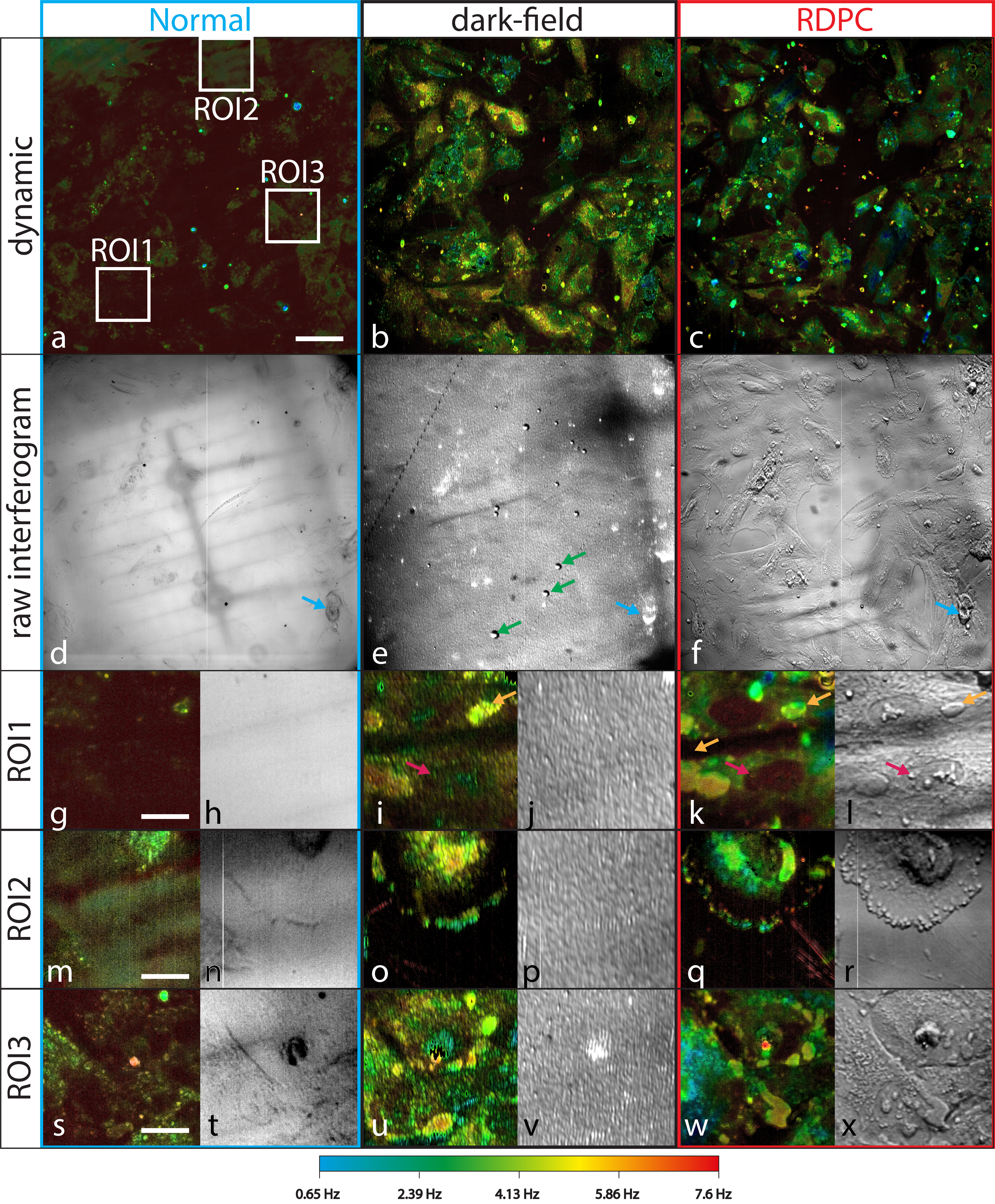}
\captionsetup{width=1.0\textwidth}
\caption{\textbf{Comparison of dynamic contrast under conventional D-FFOCT, D-dFFOCT, and D-RDPC at reflective interfaces.}
M\"uller glial cells cultured on a reflective substrate were imaged using conventional D-FFOCT (\textbf{a,d}), D-dFFOCT (\textbf{b,e}), and D-RDPC (\textbf{c,f}). Panels \textbf{a--c} show dynamic contrast images and panels \textbf{d--f} representative raw interferograms. Regions of interest highlighted in \textbf{a} are enlarged in \textbf{g--x}. Blue arrows indicate representative cellular structures visible in the raw interferograms. Orange arrows highlight localized dynamic structures observed under both dark-field FFOCT and RDPC. Dark-pink arrows indicate intracellular features visible in the enlarged regions. Green arrows indicate air bubbles present near the reference surface in the dark-field configuration. Scale bars: 60~$\mu$m (\textbf{a--f}), 20~$\mu$m (\textbf{g--l}), 18~$\mu$m (\textbf{m--r}), and 30~$\mu$m (\textbf{s--x}). Together, these observations show that both D-dFFOCT and D-RDPC recover intracellular dynamic contrast at reflective interfaces, while RDPC additionally reveals directional signatures directly visible in the raw interferograms.
}
\label{fig:comparison_modes}
\end{figure}

To investigate functional imaging at reflective interfaces, human induced pluripotent stem cell-derived M\"uller glial cells were cultured directly on mirror substrates and imaged using conventional D-FFOCT, D-dFFOCT, and D-RDPC. Conventional D-FFOCT and D-RDPC acquisitions were performed in a common-path configuration using the reflective substrate as the reference field \cite{monfort2023isr}, whereas D-dFFOCT employed a dedicated reference arm together with Fourier-plane rejection of the specular substrate reflection. Representative dynamic images and corresponding raw interferograms are shown in Fig.~\ref{fig:comparison_modes}.

Under conventional illumination, intracellular dynamic contrast remains weak despite the clear presence of adherent cells in the raw interferograms (Fig.~\ref{fig:comparison_modes}a,d). The dominant specular reflection generated by the mirror substrate overwhelms the weak fluctuations originating from intracellular dynamics, resulting in poor visualization of cellular activity \cite{Monfort2026}. These observations confirm that strong reflective interfaces constitute a major limitation for conventional dynamic interferometric imaging.

D-dFFOCT substantially restores intracellular visibility (Fig.~\ref{fig:comparison_modes}b). By rejecting the dominant specular reflection originating from the substrate, the relative contribution of intracellular scattering is enhanced, revealing cellular morphology and localized dynamic activity that remain poorly visible under conventional illumination (Fig.~\ref{fig:comparison_modes}i,o,u). These observations demonstrate that biologically relevant dynamic information remains present at reflective interfaces and can be recovered through selective rejection of the specular component.

The strongest enhancement is obtained with D-RDPC (Fig.~\ref{fig:comparison_modes}c,f). Intracellular activity becomes readily visible throughout the field of view, while the corresponding interferograms exhibit pronounced directional and contour-like signatures. Membrane boundaries, intracellular structures, and subcellular textures appear directly in the raw interferometric signal with a characteristic asymmetric contrast (Fig.~\ref{fig:comparison_modes}k,l,q,r,w,x). Importantly, many of the features highlighted by the dynamic rendering are already visible in the interferograms themselves, indicating that the enhanced contrast originates from directional information encoded during image formation rather than from temporal processing alone.

Comparison of the enlarged regions further reveals that D-dFFOCT and D-RDPC share several common dynamic features while producing qualitatively different image contrasts. Many intracellular structures become visible in both modalities, indicating that they are sensitive to related biological activity. However, the appearance of these structures differs markedly. The dark-field configuration primarily enhances localized scattering contrast through suppression of the specular background, whereas RDPC configuration additionally generates extended directional signatures, asymmetric intensity distributions, and contour-like features. These directional characteristics are largely absent from both conventional and dark-field acquisitions and are consistently observed across multiple cells and regions of interest.

Taken together, these results demonstrate that both D-dFFOCT and D-RDPC substantially improve functional imaging at reflective interfaces. While D-dFFOCT restores intracellular contrast through suppression of the dominant substrate reflection, D-RDPC reveals an additional directional contrast mechanism that is not observed under conventional illumination. The origin of this directional signal is investigated in the following sections.

\subsection{Directional dependence of static and dynamic RDPC contrast}

\begin{figure}[htbp]
\centering
\includegraphics[width=1.0\textwidth]{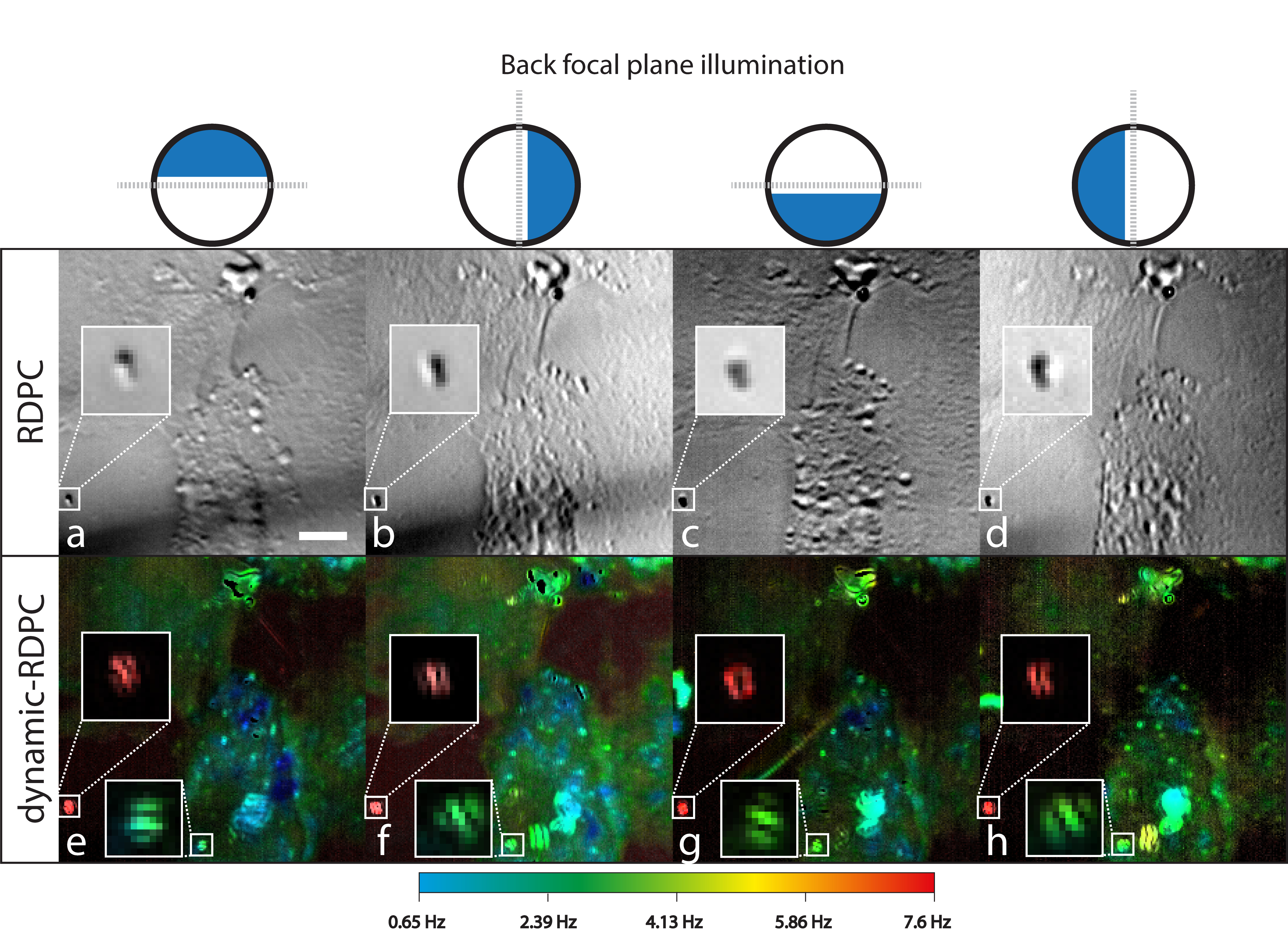}
\captionsetup{width=1.0\textwidth}
\caption{\textbf{Directional dependence of static and dynamic RDPC contrast.}
The same Müller glial cell region was imaged using four orientations of a half-pupil illumination pattern. Schematics at the top indicate the corresponding pupil geometries. Panels \textbf{a--d} show representative raw interferograms and panels \textbf{e--h} the corresponding dynamic images. Reversing the illumination direction produces pronounced contrast reversals in the interferometric signal, while rotating the illumination modifies the appearance of numerous cellular structures. These sign reversals constitute a characteristic signature of directional phase-gradient imaging. Zoom-ins highlight representative features exhibiting sign reversal upon illumination reversal. Scale bar: 15~$\mu$m.
}
\label{fig:directionality}
\end{figure}

The marked enhancement of intracellular contrast observed with D-RDPC suggests that image formation differs fundamentally from that obtained with either conventional D-FFOCT or D-dFFOCT. To investigate the origin of this contrast, the orientation of the asymmetric pupil illumination was systematically varied while imaging the same Müller glial cell region. Four illumination directions were examined, corresponding to upward, rightward, downward, and leftward half-pupil illuminations (Fig.~\ref{fig:directionality}).

The resulting interferograms (\textit{static} RDPC) exhibit a striking dependence on illumination orientation (Fig.~\ref{fig:directionality}a--d). Numerous cellular structures that appear bright for a given illumination direction become dark when the illumination is reversed by $180^\circ$. Several representative examples are highlighted by zoom-ins. In addition to these contrast reversals, rotating the illumination direction by $90^\circ$ substantially modifies the appearance of contour-like intracellular features throughout the field of view.

Such behavior cannot be explained by a simple increase in signal level or imaging sensitivity. If asymmetric illumination merely enhanced the visibility of intracellular structures, the same features would remain visible with similar contrast regardless of illumination orientation. Instead, the observed sign reversals demonstrate that the detected signal depends on the relative orientation between the illumination direction and the underlying optical structure. The measured contrast therefore behaves as a directional quantity.

The directional dependence remains visible after temporal processing (Fig.~\ref{fig:directionality}e--h). Although the dynamic images exhibit lower apparent sensitivity to illumination reversal than the raw interferograms (\textit{static} RDPC), substantial variations in contrast distribution and feature prominence remain observable. Structures that are strongly enhanced for one illumination direction often become weaker or more spatially diffuse when the illumination is rotated. The reduced magnitude of the effect compared with the \textit{static} RDPC is expected because the dynamic metric integrates temporal fluctuations over many frames and is therefore less sensitive to the sign of the underlying \textit{static} RDPC signal.

Importantly, the contrast reversals observed in the \textit{static} RDPC are characteristic of the behavior encountered in DPC imaging, where reversing the illumination direction reverses the sign of the measured signal \cite{hoffman1975modulation,mehta2009quantitative,tian2015computational}. Although the present measurements are performed in a dynamic interferometric configuration rather than a conventional phase microscope, the observed phenomenology strongly suggests sensitivity to a directional optical quantity that remains largely inaccessible under symmetric illumination.

Taken together, these observations demonstrate that RDPC does not simply improve intracellular visibility but fundamentally modifies the image-formation process. The resulting signal exhibits clear directional behavior whose sign and magnitude depend on illumination orientation, providing direct experimental evidence for a directional contrast mechanism. The influence of illumination asymmetry on the magnitude of this signal is investigated in the following section.

\subsection{Static and dynamic RDPC contrast scale with illumination asymmetry}

The directional dependence demonstrated in Fig.~\ref{fig:directionality} establishes that the observed contrast originates from a directional imaging mechanism. If this interpretation is correct, the magnitude of the signal should depend on the degree of illumination asymmetry. To test this prediction, we compared illumination geometries exhibiting different levels of pupil asymmetry.

\begin{figure}[htbp]
\centering
\includegraphics[width=1.0\textwidth]{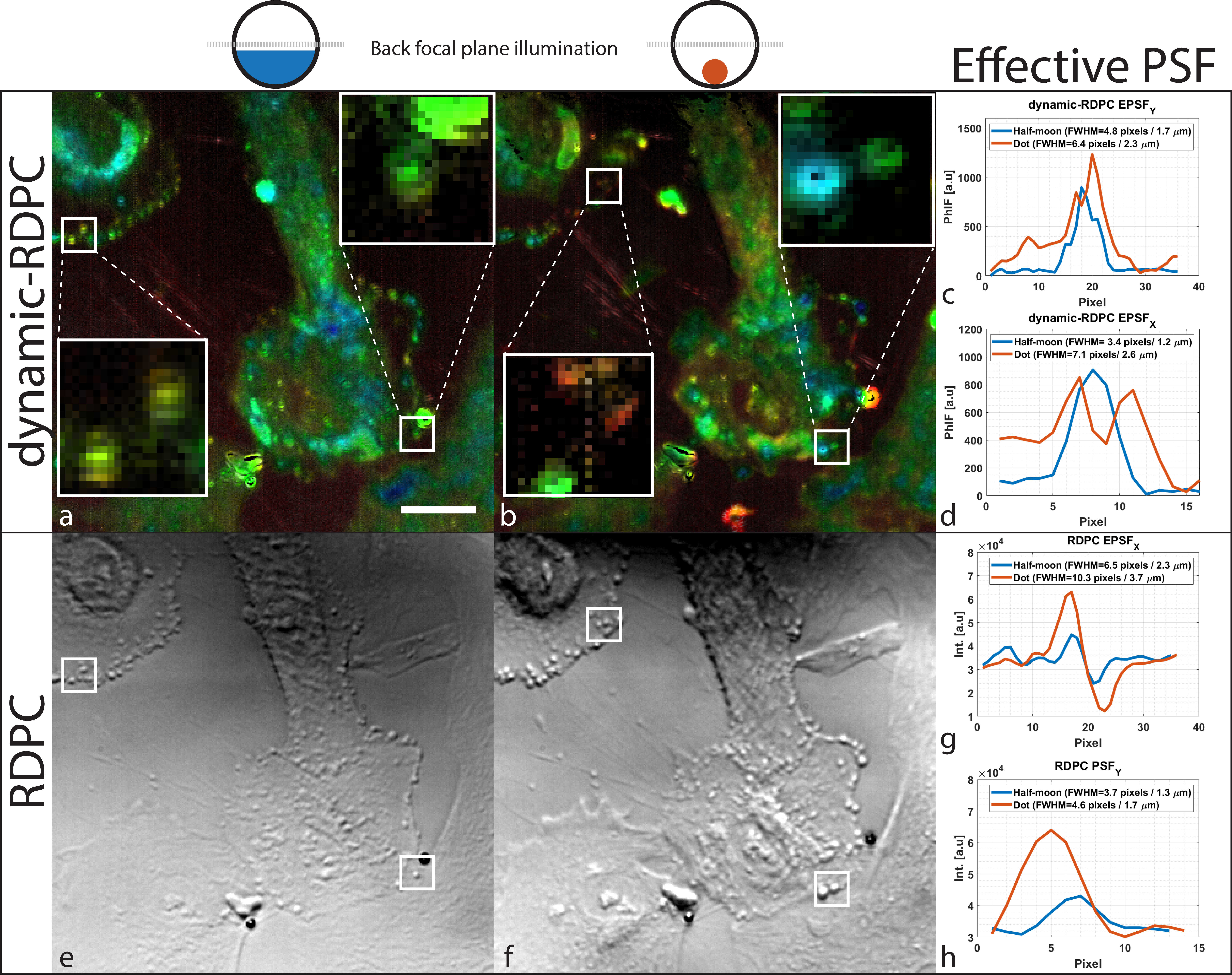}
\captionsetup{width=1.0\textwidth}
\caption{\textbf{Influence of illumination asymmetry on static and dynamic RDPC contrast.}
Comparison between half-moon and edge-dot illumination geometries. Schematics indicate the corresponding illumination distributions in the objective back focal plane. Panels \textbf{a,b} show dynamic images acquired using half-moon and edge-dot illumination, respectively. Panels \textbf{e,f} show representative raw interferograms. Profiles extracted from representative intracellular structures are shown in \textbf{c,d} and \textbf{g,h}. Increasing illumination asymmetry strengthens RDPC contrast while broadening the effective spatial response, revealing a trade-off between directional sensitivity and spatial localization. Hue scales linearly from 0.65 Hz (in blue) to 7.60 Hz (in red). Scale bar: 24~$\mu$m.
}
\label{fig:pupil_asymmetry}
\end{figure}

Two illumination geometries were compared (Fig.~\ref{fig:pupil_asymmetry}). The first consisted of a half-moon illumination pattern occupying approximately half of the objective pupil. The second employed a highly localized edge-dot illumination positioned near the edge of the objective back focal plane. Both configurations break the angular symmetry of the illumination, but the edge-dot geometry represents a substantially stronger asymmetry and selects a narrower range of illumination angles.

Both illumination geometries generate substantially stronger intracellular contrast than conventional symmetric illumination. However, the edge-dot configuration consistently produces the strongest signal throughout the field of view (Fig.~\ref{fig:pupil_asymmetry}). Cellular boundaries, intracellular structures, and localized dynamic features all become more prominent as the illumination support is reduced to a narrow angular region. Increasing illumination asymmetry therefore directly increases the magnitude of the detected dynamic contrast.

The enhancement is already apparent in the corresponding \textit{static} RDPC (Fig.~\ref{fig:pupil_asymmetry}e,f). Compared with the half-moon configuration, edge-dot illumination generates stronger contour-like signatures and larger intensity variations across intracellular structures. Features that are only weakly visible under half-moon illumination become considerably more prominent under edge-dot illumination.

This enhancement is accompanied by a progressive loss of spatial localization. Representative intensity profiles extracted from intracellular structures (Fig.~\ref{fig:pupil_asymmetry}c,d,g,h) show that edge-dot illumination produces broader spatial responses than half-moon illumination in both the \textit{static} RDPC and the corresponding D-RDPC renderings. Although the measured structures are not point objects, the systematic broadening observed throughout the field of view indicates that increased asymmetry redistributes image information over a larger spatial extent.

Importantly, the increase in contrast and the decrease in spatial localization occur simultaneously. The same illumination geometry that maximizes intracellular visibility also produces the strongest contour-like signatures and the largest apparent feature broadening. These two effects therefore appear intrinsically linked. Rather than reflecting a simple gain in signal intensity, the enhanced contrast is accompanied by a redistribution of spatial information, indicating a modification of the underlying transfer function.

Taken together, these observations reveal a clear relationship between illumination asymmetry and RDPC contrast. Increasing asymmetry systematically strengthens the directional signal identified in Fig.~\ref{fig:directionality}, while simultaneously reducing spatial localization. This behavior is consistent with the progressive dominance of an odd-symmetry transfer component as the illumination distribution departs from rotational symmetry. Combined with the contrast reversals observed upon illumination inversion, these results provide a second independent experimental signature that the detected signal originates from a directional imaging mechanism analogous to that encountered in differential phase contrast imaging.

\subsection{Directional Hilbert reconstruction recovers spatial localization}

The results of the previous section revealed a systematic trade-off between RDPC contrast and spatial localization. Increasing illumination asymmetry strengthens the directional signal but simultaneously broadens the effective spatial response of the image. Such behavior is characteristic of derivative-like imaging modalities, including differential interference contrast (DIC) and (DPC) microscopy \cite{arnison2000hilbert,arnison2004linear}. 

If the observed signal originates from a directional phase-gradient-like imaging mechanism, reconstruction approaches commonly used in DIC and DPC microscopy should partially recover the underlying object localization. To test this hypothesis, the \textit{static} RDPC images were processed using a directional Hilbert-transform (dHT) reconstruction prior to dynamic analysis.

\begin{figure}[htbp]
\centering
\includegraphics[width=1.0\textwidth]{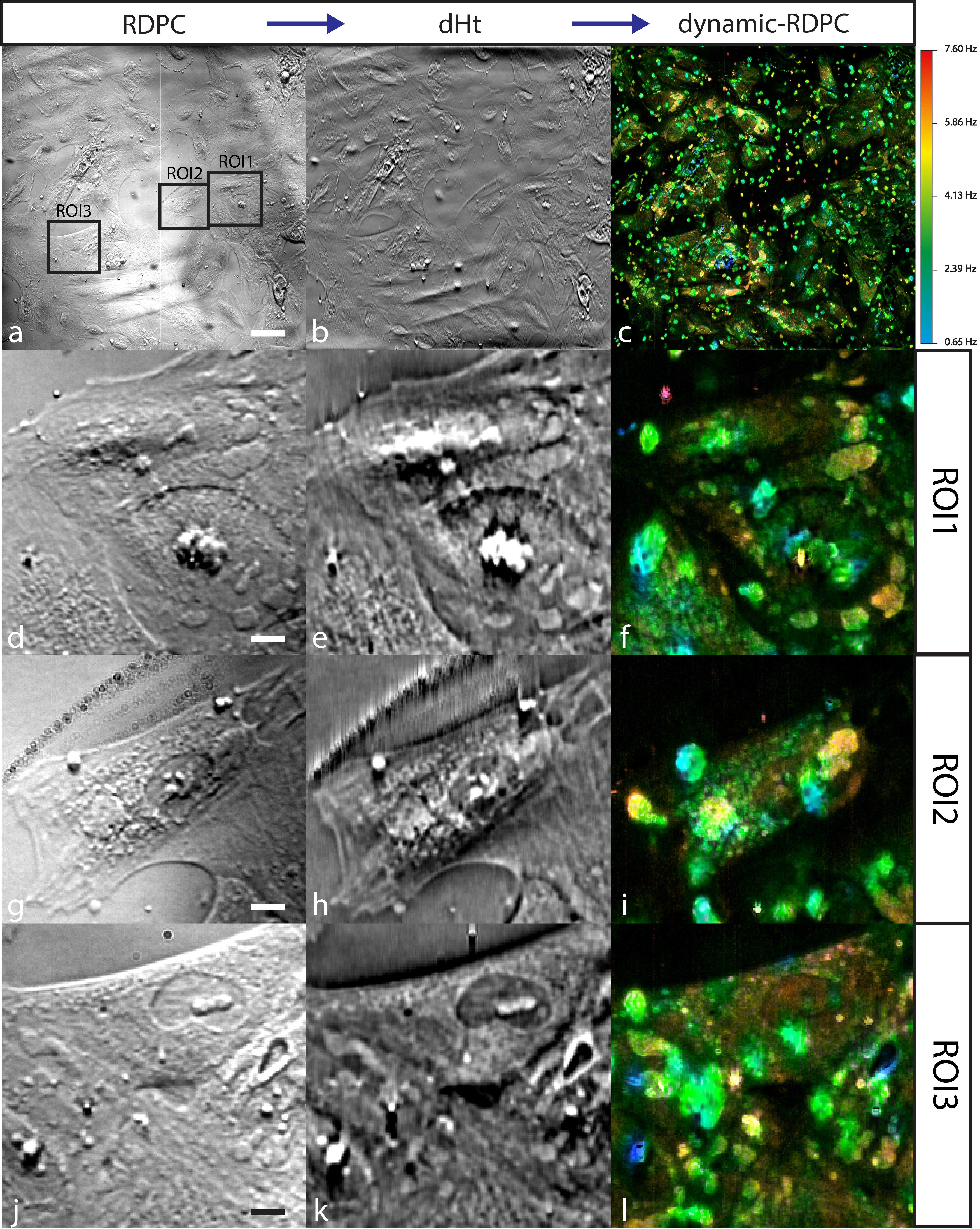}
\captionsetup{width=1.0\textwidth}
\caption{\textbf{Directional Hilbert-transform (dHT) reconstruction of RDPC interferograms.}
\textbf{a} Representative \textit{static} RDPC acquired under asymmetric illumination. \textbf{b} Corresponding \textit{static} RDPC after directional Hilbert-transform reconstruction. \textbf{c} Dynamic image calculated from the reconstructed \textit{static} RDPC sequence. Panels \textbf{d--l} show enlarged views of representative regions of interest. Reconstruction substantially improves localization and suppresses the extended directional halos characteristic of the unreconstructed RDPC signal. The same structures recovered by directional Hilbert reconstruction are subsequently visible in the corresponding D-RDPC images, demonstrating that the information responsible for the dynamic contrast is already encoded within the oblique interferograms. Scale bars are 50~$\mu$m for \textbf{a-c}, 10 $\mu$m for \textbf{d-f}, 8 $\mu$m for \textbf{g-i} and 9 $\mu$m for \textbf{j-l} 
}
\label{fig:deconvolution_workflow}
\end{figure}

The reconstruction procedure is illustrated in Fig.~\ref{fig:deconvolution_workflow}. For each interferogram, a directional Hilbert filter aligned with the illumination direction was applied in Fourier space before performing the standard D-FFOCT processing. The directional Hilbert transform has previously been shown to recover localized structures from DIC-like derivative contrast by approximately inverting the action of a directional gradient operator \cite{arnison2000hilbert,arnison2004linear}.

The reconstruction produces an immediate recovery of spatial localization (Fig.~\ref{fig:deconvolution_workflow}b). Structures that appear elongated or distributed over asymmetric halos in the \textit{static} RDPC images become more compact and spatially confined after processing. This effect is particularly evident in the enlarged regions of interest, where membrane contours, thin cellular extensions, and intracellular punctate structures become considerably easier to identify.

Importantly, the same structures subsequently appear in the dynamic images generated from the reconstructed RDPC sequences (Fig.~\ref{fig:deconvolution_workflow}c). The reconstruction therefore does not create new information. Instead, it reveals information that was already present within the \textit{static} RDPC but redistributed by the directional transfer function associated with asymmetric illumination.

\begin{figure}[htbp]
\centering
\includegraphics[width=1.0\textwidth]{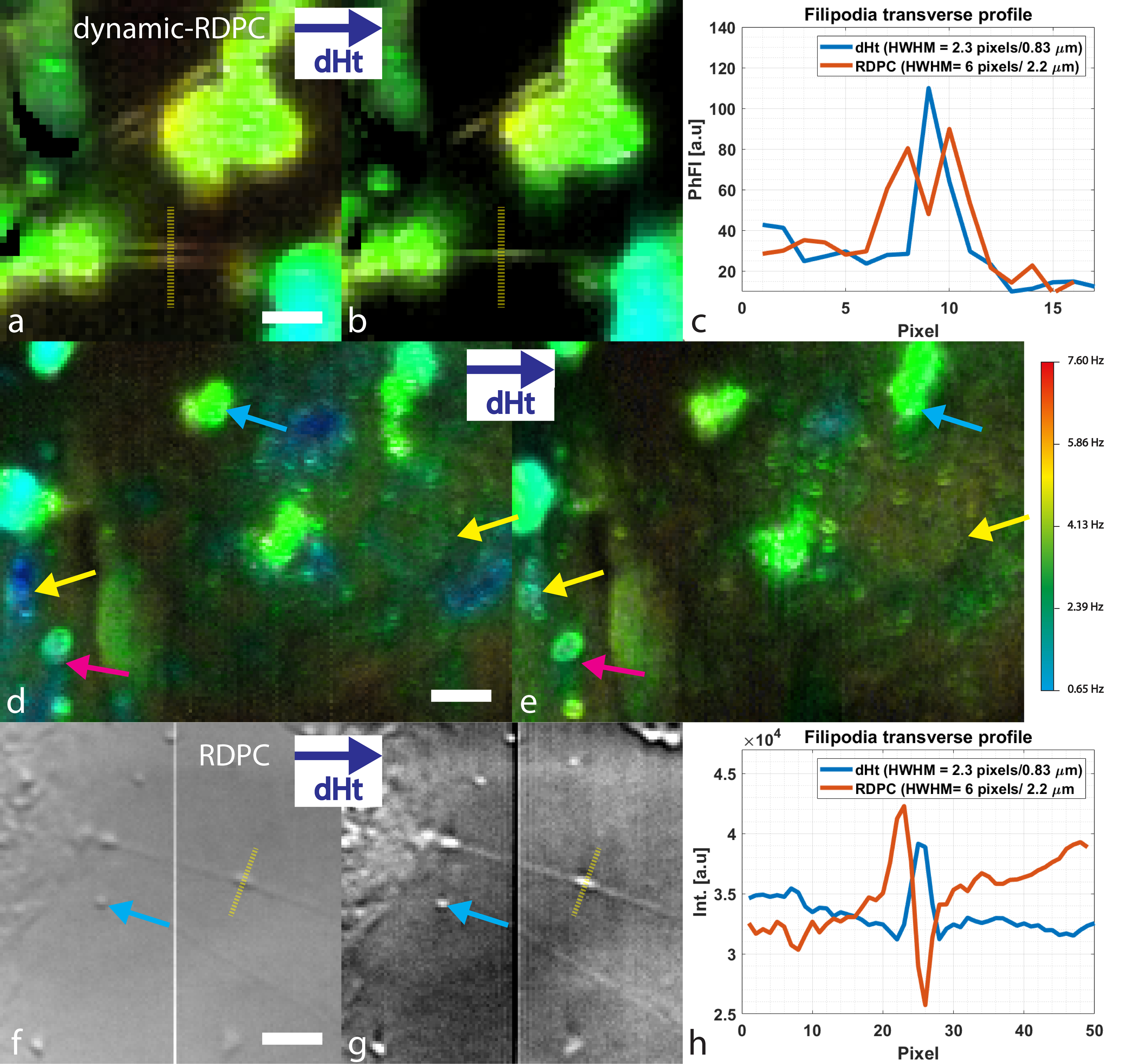}
\captionsetup{width=1.0\textwidth}
\caption{\textbf{Quantification of localization recovery following directional Hilbert-transform reconstruction.}
Representative structures are compared before and after reconstruction in both dynamic images and raw interferograms. Transverse intensity profiles demonstrate a substantial reduction in apparent feature width following reconstruction. The improved localization observed in the dynamic renderings is mirrored by a comparable narrowing in the corresponding \textit{static} RDPC, indicating that the recovered spatial information originates directly from the \textit{static} RDPC signal. These measurements demonstrate that the apparent broadening induced by strongly asymmetric illumination largely reflects a redistribution of spatial information rather than an irreversible loss of information. Yellow dashed lines mark the locations of the transverse intensity profiles used for width measurements. Yellow arrows highlight intracellular features that are only weakly visible or not readily discernible in the unreconstructed images and become clearly distinguishable following reconstruction. Pink arrows indicate the suppression of extended halo-like structures associated with the directional gradient response, resulting in improved spatial localization. Blue arrows indicate highly dynamic punctate structures, consistent with bacterial contamination, which appear as bright saturated spots in the dynamic images. Scale bar are 3 $\mu$m for \textbf{a,b} and 6 $\mu$m for \textbf{d,e,f,g}.
}
\label{fig:deconvolution_quantification}
\end{figure}

To quantify this effect, representative intracellular structures were analysed before and after reconstruction (Fig.~\ref{fig:deconvolution_quantification}).

As shown in Fig.~\ref{fig:deconvolution_quantification}, transverse profiles measured across representative thin structures exhibit a marked narrowing after reconstruction. In the examples analysed here, the apparent width decreases from approximately $2.2~\mu$m to $0.83~\mu$m. A comparable narrowing is observed directly in the corresponding \textit{static} RDPC images, demonstrating that the improved localization is not generated by the temporal processing itself but originates from information already encoded within the interferometric signal.

Beyond improving localization, the reconstruction reveals intracellular structures that are only partially discernible in the unreconstructed images. Punctate features, thin cellular extensions, and diffuse intracellular activity patterns become more clearly distinguishable following reconstruction. Importantly, these structures are not created by the reconstruction process itself. Rather, they were already encoded within the RDPC signal and become accessible once the directional transfer function is partially inverted.

Taken together, these observations demonstrate that the localization penalty associated with strongly asymmetric illumination is largely reversible. Rather than reflecting an irreversible loss of information, the broadening observed under highly asymmetric illumination arises primarily from the directional transfer function associated with RDPC image formation. The ability of a simple directional Hilbert-transform reconstruction to recover localized intracellular structures indicates that the relevant spatial information remains encoded within the \textit{static} RDPC interferograms.

The success of this reconstruction provides a third independent experimental signature supporting the interpretation that the observed contrast originates from a directional imaging mechanism closely related to differential phase contrast. Together with the contrast reversals observed upon illumination inversion and the systematic increase in signal strength with increasing illumination asymmetry, these results establish a consistent experimental framework linking the observed dynamic contrast to an underlying RDPC imaging mechanism.

\section{Discussion}

This work addresses a major limitation of functional optical imaging at highly reflective interfaces. Although reflective substrates are ubiquitous in biomedical devices, implantable materials, neural interfaces, microelectrode arrays, and engineered cell-culture platforms, the strong specular reflections generated by these surfaces often dominate the detected optical signal and substantially reduce intracellular dynamic contrast. We demonstrate that this limitation can be overcome through two complementary optical strategies: D-dFFOCT and D-RDPC.

D-dFFOCT relies on a straightforward physical principle. By suppressing the dominant specular reflection and preferentially detecting scattered light, it increases the relative contribution of intracellular fluctuations to the detected signal. Importantly, the resulting dynamic signal is equivalent to that measured in D-FFOCT for a conventional application, enabling the extensive body of work developed for D-FFOCT—including contrast interpretation, biological insights, and analytical approaches—to be directly transposed to D-dFFOCT. Consequently, D-dFFOCT reveals intracellular structures and dynamic activity that remain difficult to visualize under conventional illumination. Although dark-field implementations have previously been explored in OCT and FFOCT for structural imaging \cite{auksorius2015darkfield,xie2021darkfield,yoo2020darkfield}, the present work demonstrates, to our knowledge, the first application of dark-field filtering to D-FFOCT. These results establish D-dFFOCT as a practical solution for functional imaging at highly reflective interfaces.

The most significant finding of this work, however, concerns the directional contrast generated under asymmetric illumination. Initially implemented as a simple modification of the illumination geometry in a mirror-based imaging configuration, asymmetric illumination produced a striking enhancement of intracellular dynamic contrast together with pronounced directional signatures directly visible in the raw interferograms. Three independent observations consistently support the interpretation that this signal originates from a directional phase-gradient imaging mechanism. First, image contrast reverses when the illumination direction is reversed. Second, the magnitude of the effect increases as the illumination becomes increasingly asymmetric. Third, directional Hilbert-transform reconstruction successfully recovers spatial localization from the broadened features generated under highly asymmetric illumination. Together, these observations are characteristic of imaging systems possessing an odd-symmetry transfer function and are well-known signatures of differential phase contrast imaging \cite{hoffman1975modulation,mehta2009quantitative,tian2015computational}.

Importantly, comparison with recent reflection phase-imaging approaches suggests that the observed contrast is most consistently interpreted within a RDPC framework\cite{matlock2020inverse}. Although the present experiments were initially conceived within the framework of interface self-referenced D-FFOCT, analysis of the optical geometry indicates that the dominant signal originates from interference between a mirror-returned zeroth-order field and mirror-returned forward-scattered components. The external reference arm does not contribute to the measurements presented here. Instead, the reflective substrate simultaneously provides the reference-like field and redirects forward-scattered light back toward the objective. This optical configuration shares key features with that employed in RDPC approaches for generating \textit{static} RDPC contrast \cite{matlock2020inverse}. The principal contribution of the present work is therefore not the demonstration of \textit{static} RDPC itself, but the observation that RDPC signals exhibit biologically meaningful temporal fluctuations and can be exploited as a functional imaging modality due to its similarities with D-dFFOCT.

More broadly, the RDPC interpretation proposed here places the observed contrast within the rapidly growing field of reflection and epi-illumination phase imaging. Oblique back-illumination microscopy \cite{ford2012oblique}, reflection DPC \cite{matlock2020inverse}, epi-mode quantitative phase imaging \cite{ledwig2019epi}, and reflection quantitative phase microscopy \cite{ma2022reflection} have all demonstrated that asymmetric illumination can convert directional optical information into measurable image contrast in geometries where transmission measurements are inaccessible. The present work extends these concepts into the temporal domain by demonstrating that the directional signal itself exhibits biologically meaningful fluctuations, suggesting that temporal fluctuation analysis may be applicable across a broader class of phase-imaging modalities.

The phenomenological framework developed in Supplementary Note~1 provides a simple interpretation of these observations and extends concepts previously developed for differential phase contrast and reflection phase imaging \cite{mehta2009quantitative,tian2015qdpc,matlock2020inverse}. Within this framework, asymmetric illumination generates sensitivity to directional phase gradients through an odd-symmetry transfer function, while temporal processing extracts fluctuations of the resulting directional signal. The principal result of Supplementary Note~1 is that D-RDPC can be interpreted phenomenologically as temporal fluctuations of a directional phase-gradient signal. In its simplest form,

\begin{equation}
D(\mathbf r)
\propto
\left\langle
\left|
\Delta
\left[
\mathbf u \cdot \nabla \phi(\mathbf r,t)
\right]
\right|
\right\rangle_t ,
\end{equation}

where $\mathbf u$ denotes the illumination direction and $\phi$ the optical phase distribution. Under this interpretation, D-RDPC probes fluctuations of a directional optical quantity rather than fluctuations of a conventional interferometric scattering signal. This distinction is important because it suggests that intracellular dynamics may be interrogated through phase-gradient fluctuations that remain largely inaccessible to conventional D-FFOCT.

Beyond the mechanistic interpretation, D-RDPC may offer practical advantages for imaging weak phase objects. DPC converts local phase gradients into measurable intensity variations optically before detection \cite{mehta2009quantitative,tian2015qdpc,park2018quantitative}. This differs fundamentally from numerical differentiation performed after image acquisition. When gradients are generated numerically, detector noise is differentiated together with the signal, leading to progressive amplification of high-spatial-frequency noise. In contrast, DPC performs the gradient operation optically, before shot noise and camera read noise are added. Consequently, directional phase information is transferred to the detector with a more favourable noise transfer characteristic. This property may contribute to the strong visibility of fine intracellular structures observed throughout the present study and may be particularly advantageous for imaging weakly scattering or weakly phase-shifting biological specimens.

The stronger intracellular visibility observed in D-RDPC compared with D-dFFOCT may arise from multiple factors. First, the RDPC configuration likely derives a substantial fraction of its contrast from mirror-returned forward-scattered fields, whereas D-dFFOCT primarily relies on backward-scattered light. Since biological cells and tissues scatter predominantly in the forward direction \cite{jacques2013optical,schmitt1999optical}, D-RDPC may access a richer reservoir of intracellular optical information than dark-field interferometric detection. Second, the optical encoding of directional phase-gradient information discussed above may provide a more favorable noise-transfer characteristic than scattering-based contrast alone \cite{mehta2009quantitative,tian2015qdpc,park2018quantitative}. The combination of forward-scattering sensitivity and optical gradient encoding may therefore contribute to the particularly strong intracellular visibility observed in D-RDPC. Further work will be required to determine the relative importance of these mechanisms.

The present implementation should nevertheless be regarded as a proof of principle rather than an optimized realization of D-RDPC. The edge-dot illumination geometry used here increases directional sensitivity at the cost of photon efficiency because only a small fraction of the available illumination fills the objective pupil. Previous work in quantitative differential phase contrast microscopy has demonstrated that illumination distributions can be optimized to maximize phase-gradient transfer while maintaining a large pupil filling factor \cite{tian2015qdpc,fan2019optimal}. Semi-annular, segmented, or continuously varying asymmetric pupil illuminations may therefore provide substantially stronger directional contrast than the simple edge-dot geometry employed here while preserving a significantly larger photon budget. At the same time, although strongly asymmetric illumination enhances directional contrast, it also broadens the effective spatial response and reduces apparent localization. The success of directional Hilbert-transform reconstruction demonstrates that much of this degradation reflects a redistribution of information rather than a true loss of information. Future implementations could therefore combine optimized pupil engineering with reconstruction approaches derived from quantitative DPC and Fourier ptychography \cite{zheng2013fourier,tian2015qdpc,fan2019optimal}. Multi-angle D-RDPC acquisitions implemented through rotating illumination patterns, scanning mirrors, or digital micromirror devices could potentially exploit concepts developed in quantitative DPC and Fourier ptychography to recover additional spatial-frequency content while preserving directional sensitivity \cite{zheng2013fourier,tian2015qdpc,fan2019optimal}.

The mirror-based implementation described here also opens opportunities that are not readily accessible in conventional DPC microscopy. Most notably, the use of low-coherence interferometric detection provides a natural route toward optical sectioning. Combining directional phase-gradient imaging with coherence gating could ultimately enable D-RDPC measurements with depth selectivity, extending phase-gradient imaging toward three-dimensional biological systems while retaining sensitivity to intracellular dynamics. In this respect, D-RDPC may provide a bridge between the fields of phase-gradient microscopy and coherence-gated imaging \cite{hillmann2016holoscopy,drexler2008oct}, combining optical sectioning with sensitivity to directional phase-gradient fluctuations.

Perhaps the most exciting implication of this work is that DPC microscopy itself may be extended from a structural modality toward a functional one. Conventional DPC is now widely available on commercial microscopes and routinely used for label-free imaging of living cells \cite{hoffman1975modulation,mehta2009quantitative,tian2015computational}. The observation that directional phase-gradient signals contain measurable temporal fluctuations raises the possibility of dynamic DPC (D-DPC) imaging, in which endogenous intracellular activity is extracted directly from phase-gradient dynamics. In this sense, D-DPC may play for DPC microscopy a role analogous to that played by D-FFOCT for conventional FF-OCT: transforming a structural imaging modality into a functional one through temporal fluctuation analysis. If confirmed, such an approach could provide a simple and broadly accessible route toward functional label-free microscopy, particularly because DPC hardware is already integrated into many commercial microscope platforms.

\section*{Conclusion}

In this work, we investigated dynamic imaging at highly reflective interfaces and identified two complementary strategies for recovering intracellular dynamic contrast in the presence of strong specular reflections.

First, we introduced D-dFFOCT, which suppresses the dominant substrate reflection and restores intracellular dynamic contrast through selective detection of scattered light. This approach provides a practical solution for functional imaging on reflective substrates where conventional D-FFOCT exhibits limited sensitivity.

Second, we demonstrated that asymmetric illumination generates a distinct form of directional dynamic contrast. Through systematic investigation of illumination direction, pupil asymmetry, and directional Hilbert-transform reconstruction, we showed that the observed signal exhibits the characteristic signatures of a DPC imaging mechanism. Comparison with reflection phase-imaging approaches and the phenomenological framework developed in Supplementary Note~1 indicate that this contrast is most consistently interpreted as D-RDPC.

The principal contribution of this work is therefore not the demonstration of \textit{static} RDPC itself, but the observation that RDPC signals support temporal fluctuation analysis and can be exploited as a functional imaging modality. More broadly, these findings suggest that DPC microscopy may be extended from a structural modality toward a functional one through temporal fluctuation analysis. By enabling endogenous contrast from phase-gradient dynamics, D-RDPC establishes a new form of label-free functional imaging that complements existing interferometric approaches while remaining closely connected to widely available DPC microscopy platforms. Beyond the specific mirror-based implementation demonstrated here, this framework may open new opportunities for investigating intracellular dynamics, cell--surface interactions, and biomaterial--tissue interfaces, while providing a broadly accessible route toward functional phase-gradient microscopy.

\section{Methods}

\subsection{Optical setup}

\begin{figure}[htbp]
\centering
\includegraphics[width=1.0\textwidth]{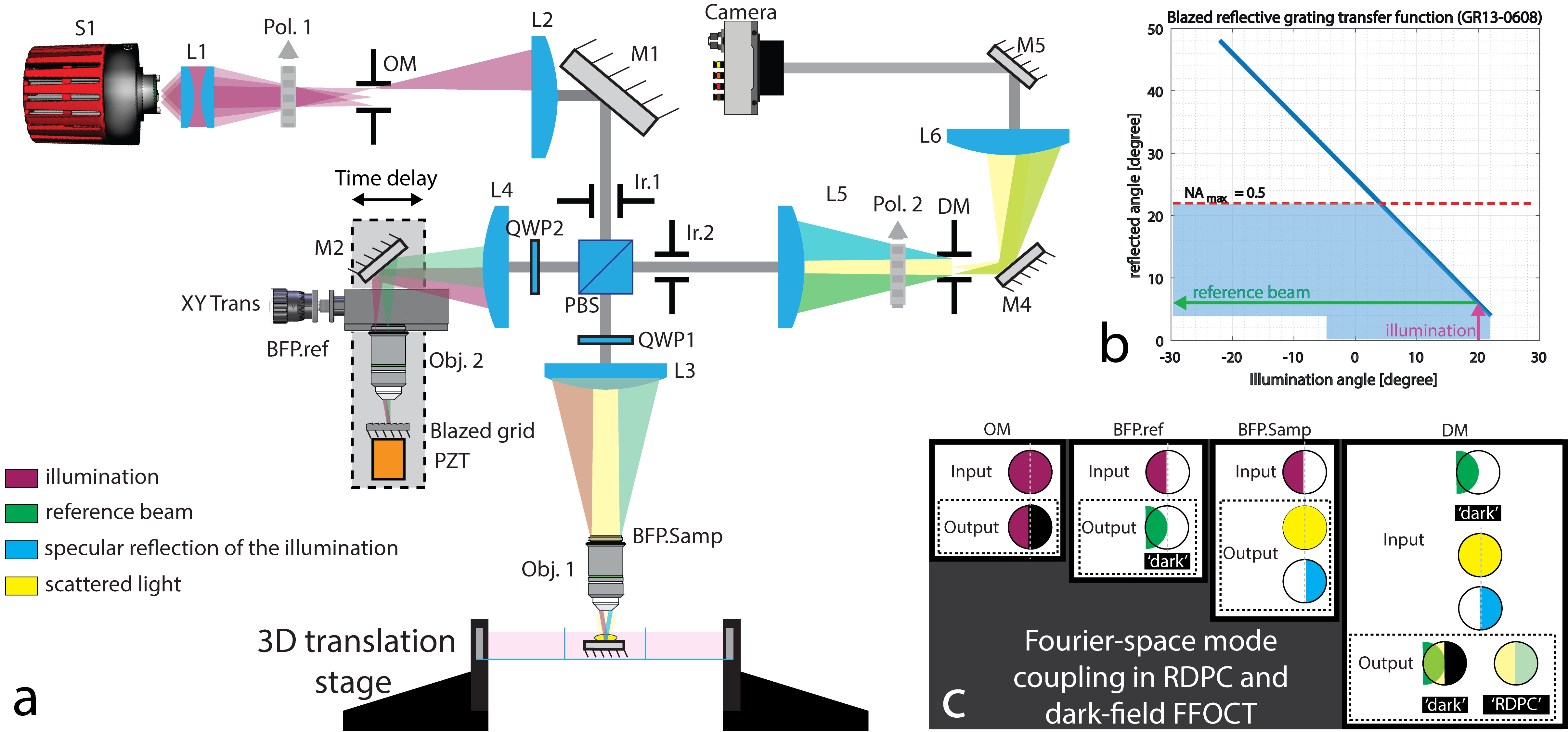}
\captionsetup{width=1.0\textwidth}
\caption{\textbf{Experimental implementation of D-RDPC and D-dFFOCT.} (\textbf{a}) Optical setup. Illumination from a broadband LED source is relayed onto an illumination mask (OM) conjugated to the back focal planes of the microscope objectives. The system is based on a modified Linnik interferometer. For D-dFFOCT, a dedicated reference arm is implemented using a blazed reflective grating placed at the reference position. For D-RDPC, the dedicated reference arm is removed and the specular reflection from the reflective substrate supporting the biological sample serves as the reference field in a common-path geometry. A Fourier-plane mask (DM), conjugated to the objective pupils, enables selective spatial-frequency filtering. The sample is mounted on a motorized three-dimensional translation stage. (\textbf{b}) Angular transfer function of the blazed reflective grating (GR13-0608). The grating introduces an angular offset between the illumination and reference modes. The shaded region indicates the angular acceptance of the objective ($\mathrm{NA}=0.5$). The grating period was selected such that the reference field remains within the collection numerical aperture while the dominant specular reflection can be displaced outside the Fourier-space region transmitted by the dark-field mask. (\textbf{c}) Fourier-space mode coupling associated with D-dFFOCT and D-RDPC. The illumination mask (OM) defines the angular illumination distribution. In the dark-field configuration, the grating shifts the reference-arm spatial modes relative to the illumination modes, allowing the specular reflection to be rejected while preserving interference between the scattered sample field and the reference field. In the D-RDPC configuration, asymmetric illumination selectively excites one side of the pupil, breaking the symmetry between opposite spatial frequencies and generating directional phase-gradient contrast. The diagrams illustrate the evolution of the illumination, reference, specular, and scattered-field distributions through the conjugated pupil planes.
}
\label{fig:setup}
\end{figure}

The experimental setup is shown in Fig.~\ref{fig:setup}a. Imaging was performed using a custom FF-OCT microscope based on a modified Linnik interferometer architecture. Illumination was provided by a light-emitting diode (M810L5, Thorlabs, USA) centered at 810~nm with a spectral bandwidth of 30~nm. The LED emitter acted as an extended spatially incoherent source and was relayed onto an illumination mask (OM) using a pair of achromatic lenses (L1). The illumination mask was positioned in a plane conjugated to the back focal planes of the microscope objectives, allowing controlled shaping of the illumination distribution within the objective pupils.

The illumination beam was directed toward the microscope through a polarizing beam splitter (PBS). Two water-immersion objectives (UMPLFLN20XW, Olympus/Evident, Japan; NA = 0.5) were arranged in a Linnik configuration. One objective illuminated and collected light from the sample, while the second objective formed a conjugate arm used for interferometric detection.

Quarter-wave plates were inserted in both interferometer arms. Following double passage through the quarter-wave plates, the reflected fields underwent a $90^\circ$ polarization rotation, enabling efficient recombination through the PBS. A linear polarizer (Pol.~2) positioned before the camera projected the fields onto a common polarization state prior to detection.

The optical system was designed such that the illumination mask (OM), the objective back focal planes (BFP.Samp and BFP.Ref), and the Fourier filtering plane (DM) were mutually conjugated. This arrangement enabled independent control of the illumination angular distribution and Fourier-space filtering. Interferometric images were recorded using a CMOS camera (Q-2HFW, Adimec, Netherlands) through relay optics (L5--L6).

Samples were mounted on a motorized three-axis translation stage (RAMM RM-1250 and LS-100, ASI, USA), allowing precise positioning and focus adjustment during image acquisition.

\subsection{D-RDPC and D-dFFOCT implementations}

Two distinct imaging modalities were implemented using the optical platform described above: D-RDPC and D-dFFOCT.

For D-RDPC imaging, the dedicated reference arm was disabled by adjusting the polarization state such that only the sample arm contributed to the detected signal. In this configuration, the reflective substrate supporting the biological sample simultaneously provided the mirror-returned zeroth-order field and redirected a portion of the forward-scattered sample field back toward the objective. Image formation therefore relied on interference between the mirror-returned field and the sample-scattered field in a common-path geometry.

Directional contrast was generated by introducing asymmetric illumination distributions in the illumination pupil (OM). Half-moon and edge-dot illumination geometries were employed throughout this study, producing controlled asymmetry within the objective back focal plane. Representative Fourier-space mode distributions associated with these illumination patterns are shown schematically in Fig.~\ref{fig:setup}c.

For D-dFFOCT, a dedicated reference arm operating in a conventional Linnik interferometer configuration was employed. A blazed reflective grating (GR13-0608, Thorlabs, USA) replaced the planar reference mirror typically used in FFOCT systems. As illustrated in Fig.~\ref{fig:setup}b, the grating introduced an angular offset between the reference field and the illumination field. This offset allowed the dominant specular reflection originating from the reflective substrate to be rejected using a Fourier-plane stop (DM), while preserving interference between the scattered sample field and the reference field.

As a result, only scattered sample light that did not overlap with the specular reflection in Fourier space contributed significantly to the detected interferometric signal. This configuration strongly suppressed the substrate reflection while enhancing the relative contribution of intracellular scattering. A half-moon illumination pattern was used throughout the dark-field experiments. The corresponding Fourier-space filtering process is illustrated in Fig.~\ref{fig:setup}c, which shows the evolution of the illumination, reference, specular, and scattered-field distributions through the successive conjugate pupil planes of the system.

Compared with previously reported dark-field FFOCT implementations \cite{auksorius2015darkfield}, the present configuration fills the objective pupil asymmetrically along a single dimension. In principle, this geometry preserves the full spatial-frequency support along the orthogonal dimension while enabling efficient rejection of the specular substrate reflection.

The Fourier-space mode coupling associated with both D-RDPC and D-dFFOCT is summarized schematically in Fig.~\ref{fig:setup}c.

\subsection{Müller glial cell culture}

Human induced pluripotent stem cell-derived Müller glial cells were generated and maintained as previously described \cite{couturier2021muller}. Cells were cultured in DMEM/F12 medium (Thermo Fisher Scientific, 11320074) supplemented with 10\% fetal bovine serum (Thermo Fisher Scientific, A5256701) and maintained under standard culture conditions (37$^\circ$C, 5\% CO$_2$). Culture medium was renewed three times per week.

For imaging experiments, cells were seeded directly onto silver-coated reflective substrates. Prior to cell seeding, substrates were coated with Geltrex\textsuperscript{\textregistered} LDEV-Free, hESC-Qualified, Reduced Growth Factor Basement Membrane Matrix (Thermo Fisher Scientific, A1413302). Briefly, a 1:100 dilution of Geltrex in cold DPBS was applied to the substrate surface and incubated for 1~h at 37$^\circ$C before cell plating.

All imaging experiments were performed in phenol-red-free culture medium at room temperature.

\subsection{Image acquisition}

For all imaging modalities, interferometric image sequences were recorded and subsequently processed to generate dynamic images from the temporal fluctuations of the detected signal. Unless otherwise stated, comparisons between imaging modalities were performed on identical sample regions using identical temporal processing parameters.

For each acquisition, a sequence of 512 interferograms was recorded at a frame rate of 50~frames~s$^{-1}$, corresponding to a total acquisition time of 10.24~s. Exposure times were adjusted according to the illumination geometry and imaging modality to maximize the use of the camera dynamic range while avoiding saturation. Exposure times were 1~ms for symmetric illumination, 5~ms for half-moon RDPC, 12~ms for edge-dot RDPC, and 19~ms for half-moon dark-field FFOCT.

Dynamic images were subsequently computed from the temporal fluctuations of the interferometric intensity at each pixel using the processing framework described below.

Unless otherwise stated, all images and quantitative analyses presented in this work were generated from a single 512-frame acquisition sequence.

\subsection{Dynamic image computation}

Dynamic images were computed from temporal fluctuations of the recorded interferometric intensity using the dynamic full-field optical coherence tomography (D-FFOCT) processing framework previously described in Refs.~\cite{Monfort2026,scholler2020dynamic}.

The image sequence was first normalized by its spatially averaged intensity at each time point in order to compensate for temporal illumination fluctuations. Dynamic contrasts were then extracted from the temporal evolution of the interferometric signal at each pixel.

The brightness channel corresponds to the Phase Fluctuation Index (PhFI), recently introduced in Ref.~\cite{Monfort2026},

\begin{equation}
\mathrm{PhFI}(\mathbf r)=
\left\langle
\left|
I(\mathbf r,t+\Delta t)-
I(\mathbf r,t)
\right|
\right\rangle_t,
\end{equation}

where $I(\mathbf r,t)$ denotes the interferometric intensity at position $\mathbf r$ and time $t$. The PhFI quantifies the average magnitude of temporal fluctuations and provides a robust estimate of local dynamic activity.

To characterize the temporal frequency content of these fluctuations, the temporal Fourier transform was computed independently at each pixel. Following temporal binning by a factor of four, the normalized magnitude spectrum was used to calculate its first and second statistical moments, corresponding respectively to the mean fluctuation frequency and the standard deviation of the fluctuation frequency distribution.

For visualization, the mean frequency was mapped to \textit{hue}, the standard deviation frequency to \textit{saturation}, and the Phase Fluctuation Index to \textit{brightness}, generating the final Hue--Saturation--Brightness (HSB) rendering used throughout this work. A working implementation of the dynamic Hue--Saturation--Brightness (HSB) rendering used in this work is publicly available at:
\url{https://github.com/noahheldt/A-guide-to-dynamic-OCT-data-analysis/blob/main/HSB%20visualization%20of%20PSD%20moments/doct_MSm_hsv_gpu_compression4_monfort.m}

\subsection{Directional Hilbert-transform reconstruction}

Directional reconstruction was applied to the RDPC interferometric image sequences acquired under asymmetric illumination. The reconstruction was inspired by Hilbert-transform approaches previously developed for differential interference contrast and differential phase contrast microscopy \cite{arnison2000hilbert,arnison2004linear}, where the measured image can be interpreted as the action of a directional derivative operator on the underlying object.

Each interferogram was first converted to double precision and centered by subtracting its median intensity value. A directional Fourier-domain reconstruction was then applied independently to each frame in order to partially invert the directional transfer function associated with asymmetric illumination.

For an image $g(x,y)$ and an illumination direction described by the unit vector $\mathbf u$, the Fourier transform $G(\mathbf k)$ was multiplied by the regularized directional integration filter

\begin{equation}
R_{\mathbf u}(\mathbf k)=
\frac{-i(\mathbf k\cdot\mathbf u)}
{(\mathbf k\cdot\mathbf u)^2+\epsilon^2},
\end{equation}

where $\mathbf k=(k_x,k_y)$ denotes the transverse spatial-frequency vector and $\epsilon$ is a regularization parameter introduced to avoid divergence near $\mathbf k\cdot\mathbf u=0$. Unless otherwise stated, $\epsilon=0.02$ was used throughout this work.

The reconstructed image was obtained as

\begin{equation}
g_{\mathrm{rec}}(x,y)=
\mathrm{Re}
\left\{
\mathcal{F}^{-1}
\left[
G(\mathbf k)
R_{\mathbf u}(\mathbf k)
\right]
\right\},
\end{equation}

where $\mathcal{F}^{-1}$ denotes the inverse Fourier transform.

For the datasets presented in Figs.~\ref{fig:deconvolution_workflow} and \ref{fig:deconvolution_quantification}, the reconstruction direction was chosen to coincide with the illumination asymmetry axis. In the case of vertical half-moon illumination, this corresponds to $\mathbf k\cdot\mathbf u=k_y$.

A slowly varying background was removed by subtracting a Gaussian-smoothed version of the reconstructed image. The reconstruction was applied independently to all interferograms prior to dynamic image computation.

A working implementation of the Hilbert-transform reconstruction used in this work is publicly available at:
\url{https://github.com/Tual29/Directional-Hilbert-transform-reconstruction}

\begin{backmatter}

\bmsection{Funding}

This work was supported by IHU FOReSIGHT through the RETINA-LENS project (P-RETLEN-IHU-000) and by the French National Research Agency (ANR) through the No-D-EYE project (M25JRAR029).

\bmsection{Acknowledgments}

The author warmly thanks Camille Hourton and Sacha Reichman for the preparation of the Müller glial cell cultures used in this study.

The author thanks Stéphane Fouquet and Gilles Tessier for kindly providing the microscope objectives used in this work.

The author is particularly grateful to Kate Grieve for her continuous support throughout this project.

\bmsection{Data Availability Statement}

The data supporting the findings of this study are available from the corresponding author upon reasonable request.

\bmsection{Supplemental document}

See Supplementary Note~1 for supporting information, including the phenomenological framework for dynamic reflection differential phase contrast and a comparison with Dynamic Optical Transmission Tomography (DOTT).

\bmsection{Disclosures}

The author declares no conflicts of interest.

\end{backmatter}

\end{document}